\documentclass[aps, prm, twocolumn, floatfix, superscriptaddress, reprint, 10pt]{revtex4-2}  
\usepackage{bm}
\usepackage{graphicx}
\usepackage[usenames]{color}
\usepackage{microtype}
\usepackage{amsmath}
\usepackage{amssymb}
\usepackage{xspace}
\usepackage{footnote}
\usepackage{relsize}
\usepackage{hhline}
\usepackage{hyperref}

\newcommand{\br}{\boldsymbol{r}}
\newcommand{\be}{\begin{equation}}
\newcommand{\ee}{\end{equation}}

\usepackage{floatrow}
\newfloatcommand{capbtabbox}{table}[][\FBwidth]

\usepackage[normalem]{ulem}
\definecolor{tangerine}{rgb}{0.944,0.522,0}
\definecolor{verde}{rgb}{0.,0.6,0}
\definecolor{rosso}{rgb}{0.9,0.0,0.2}
\definecolor{magenta}{rgb}{0.9,0.2,0.9}

\newif\ifhighlight

\newcommand{\highlight}{\highlighttrue}
\highlight
\newcommand{\editor}[2]{%
  \expandafter\newcommand\csname #1note\endcsname[1]{%
    \textcolor{#2}{(\textbf{#1:} ##1)}}%
  \expandafter\newcommand\csname #1\endcsname[1]{%
    \ifhighlight\textcolor{#2}{##1} \else ##1\fi}%
  \expandafter\newcommand\csname #1cancel\endcsname[1]{%
    \ifhighlight\textcolor{#2}{\sout{##1}}\fi}%
  \expandafter\newcommand\csname #1change\endcsname[2]{%
    \ifhighlight\textcolor{#2}{\sout{##1} ##2}\else ##2\fi}%
  \newenvironment{#1text}{\ifhighlight\color{#2}\fi}{\color{black}}
}

\editor{AG}{verde}
\editor{FG}{rosso}
\editor{resub}{blue}

\begin{document}

\title{
Effect of a Temperature Gradient on the Screening Properties of Ionic Fluids
}

\author{Andrea Grisafi}
\email{andrea.grisafi@ens.psl.eu}
\affiliation{PASTEUR, D\'epartement de chimie, \'Ecole Normale Sup\'erieure, PSL University, Sorbonne Universit\'e, CNRS, 75005 Paris, France}

\author{Federico Grasselli}
\email{federico.grasselli@epfl.ch}
\affiliation{Laboratory of Computational Science and Modeling, IMX, \'Ecole Polytechnique F\'ed\'erale de Lausanne, 1015 Lausanne, Switzerland}

\begin{abstract}
 The electrostatic screening properties of ionic fluids are of paramount importance in countless physical processes. 
 {Yet, the  behavior of ionic conductors out of thermal equilibrium has to date mainly been studied in the context of thermodiffusion phenomena by virtue of direct extensions of Debye-H\"uckel theories. We investigate how the static response of a symmetric ionic fluid is influenced by the presence of a thermal gradient by introducing a theory of electrostatic screening under a stationary temperature profile.} By borrowing mathematical methods commonly used in the semiclassical approximation of quantum particles, we find analytical solutions to the asymptotic decay of the charge density which can be used to describe the non-equilibrium response of the system to external charge perturbations {and for arbitrary ionic concentrations.} Notably, a transition between monotonic and oscillatory screening regimes is observed as an effect of the temperature variation which generalizes 
 {known results of thermal equilibrium to out of equilibrium conditions. A final quantitative example on the screening of charged surfaces in aqueous electrolytes shows that the deviation from thermal equilibrium predicted by our solutions is generally larger than thermodiffusion effects, and should therefore be taken into account for a comprehensive description of the electrical double layer.} 
 Our findings pave the way to the rigorous treatment of non-equilibrium steady states in ionic systems with potential applications to the study of energy materials, {nanostructured systems} and waste-heat-recovery technologies.
\end{abstract}

\maketitle

\section{Introduction}
Many physical and technological applications which involve ionic systems require to deal with the coexistence of electric fields and thermal gradients. Temperature gradients naturally arise in electrochemical devices as a consequence of the Joule effect~\cite{Lin1982,*frankel1982reply,Shi2022} and the heat released during electrochemical reactions~\cite{Guillamon2021}, and they are responsible for the creation of a net ionic current in thermoelectric cells~\cite{borset2015}. {A paradigmatic playground, of compelling technological relevance in opto- and micro-fluidics~\cite{Liu2021opto,Cha2022,Tian2022}, nanodevices and nanoengines~\cite{Ding2022}, and microbiology~\cite{Liu2019}, is that of charged micro-/nano-particles (e.g., metal-capped colloids) dissolved in a salt solution, which are heated selectively by laser irradiation. This leads to the onset of intense microscopic thermal gradients ($\sim 10\, \mathrm{K} / \mu \mathrm{m}$), coexisting with electric fields, and dictating particles' dynamics~\cite{Ly2018}. In turn, controlling micro- and nano-particle equilibrium and non-equilibrium arrangements offers a route to fine-tune their chemophysical properties~\cite{Rossi2020,Rossi2022CopperNanocatal}.}
Furthermore, ionic fluids are playing an increasingly prominent role as heat-transfer fluids in solar energy storage~\cite{Reddy2011,Minea2020} and waste-heat-recovery technologies~\cite{Gibbs2019,Kjelstrup2023}, as well as in the design of sustainable molten salt reactors for future nuclear energy systems~\cite{Wang2018}.  {An accurate description of all these phenomena requires a deep understanding of how the electrostatic screening properties of an ionic conductor are affected by the presence of an inhomogeneous temperature distribution.}

{Recently, there has been a growing effort in the modeling of thermal transport properties of ionic fluids \cite{Ohtori2009,Bertossa2019,Grasselli2021}
as well as in the description of thermodiffusion effects such as Seebeck and Soret~\cite{Vos2022,Wurger2020,Drigo2023}. Beyond these phenomena, most of theoretical studies that explicitly treated the effect of a temperature gradient on the static screening response of ionic fluids have involved more or less direct extensions of Debye-H\"uckel (DH) theories~\cite{Zhang-Zhao2019,Dhur2006pnas,Dhont2008,Liu2021}. One of the major difficulties consists in providing a formal microscopic generalization of thermodynamic laws and related variational principles to non-isothermal situations~\cite{Schmidt2011,Wittkowski2012,Anero2013,Hsieh2020}. For instance, liquid-vapor interfaces have been investigated by making use of temperature-dependent influence parameters that are derived from experimentally parametrized equations of state~\cite{Bedeaux2003,Johannessen2003,Magnanelli2014}. However, the lack of experimental data implies that similar approaches can be hardly applied to the study of ionic fluids, especially for those systems that involve large ionic concentrations.}

{When it comes to thermal equilibrium, the theory of liquids which describes the static charge-density response in ionic fluids} is well established and predicts the capability of the system to perfectly screen an external charge over a finite length scale~\cite{Hansen_McDonald,Stillinger_Lovett,Perkin2016}. The electrostatic screening is manifested either as a pure exponential decay of the fluid charge density distribution~\cite{Debye_Hueckel} or as an exponentially damped oscillation that reflects a shell-like ordering of the charge carriers~\cite{Ballone1981first,Ballone1981second,carvalho1994,Rotenberg2018,Coupette2018}.
A crossover between a monotonic and oscillatory behavior, in particular, is observed at decreasing temperatures and/or increasing ionic concentrations and takes the name of Fisher-Widom line~\cite{fisher-widom}. 
{By and large, understanding how a similar description can be extended to non-equilibrium steady states represents a goal of particular interest.}

In this article, we provide a microscopic  theory of electrostatic screening in ionic fluids under a linear and stationary temperature profile. In particular, we show that the Wentzel-Kramers-Brillouin (WKB) method, commonly applied to quantum systems, can be adopted to find analytical solutions to the decay of the charge density, where the temperature gradient plays the role of $\hbar$ in the semiclassical asymptotic expansion. 
In so doing, we generalize the transition between monotonic and oscillatory screening regimes under steady-state conditions, recovering the thermal equilibrium solutions as a special limit. {Explicit solutions are given for the screening of a planar surface charge, assumed as the origin of the external perturbation. We finally apply our theory to quantitatively estimate the effect of a thermal gradient on the screening charge distribution of typical electrolytic solutions, showing that the effect is larger than that induced by the ionic Seebeck effect.}

\section{General theory}
{Let us consider a classical Coulomb fluid made of two ionic species of opposite charge , $Z_+=-Z_-=Z$.
Unless otherwise specified, we will adopt Hartree atomic units. The particles' pair-potential is defined as
$u_{ij}(r) = u_{ij}^{SR}(r) + Z_iZ_j/r$, where $u_{ij}^{SR}(r)$ represents the short-range ionic interaction. 
We are interested in the screening response of the charge density distribution $\rho_Q(\br)$ of the ionic fluid to an external charge, under an inhomogeneous temperature profile~$T(\br)$.
We assume the fluid is at local thermodynamic equilibrium, so that the local chemical potential of the two ions $\mu_{+/-}(\br)$, can be defined~\cite{mazur1984,Dhur2006}.
For convenience, we will adopt a unitary transformation of the ionic variables which allows us to formulate the theory in terms of total charge ($Q$) and particle number ($N$) linear combinations. For example, the chemical potentials of the total charge and particle number are $\mu_{Q}(\br) = Z [\mu_+(\br) - \mu_-(\br)]$ and $\mu_{N}(\br) = \mu_+(\br) + \mu_-(\br)$, respectively. An analogous transformation allows us to define the charge and number densities, $\rho_Q(\br)$ and $\rho_N(\br)$.}

\subsection{Non-isothermal density functional theory}

We rely on density-functional theory for classical particles, a.k.a., classical-DFT~\cite{Mermin,Evans2016}, generalized for the inclusion of a stationary temperature distribution~\cite{Anero2013,Wittkowski2012}. 
In particular, we rely on the result of Ref.~\cite{Anero2013}, which provides a generalization of classical-DFT to non-isothermal conditions. In this framework, the stationary state is characterized by the maximization of a non-equilibrium entropy functional with respect to the charge density $\rho_Q(\br)$, total number density~$\rho_N(\br)$ and energy density~$e(\br)$:
\begin{equation}
\begin{split}
    &S_\text{NE}[e,\rho_N,\rho_Q] = -\tilde{F}_\mathrm{NE}[T, \rho_N, \rho_Q] + \int d\br\,\frac{e(\br)}{T(\br)} .
\end{split}
\end{equation}
Here, $\tilde{F}_\mathrm{NE}$ is the non-isothermal extension of the (dimensionless) Helmholtz free energy~\cite{Anero2013}. In general, each of $\mu_Q(\br)$, $\mu_N(\br)$ and $T(\br)$ is a functional of $\rho_Q(\br)$, $\rho_N(\br)$ and $e(\br)$, and the stationary state is characterized by the solution of three coupled Euler-Lagrange equations \footnote{Following Ref. \cite{Anero2013}, $\tilde{F}_\mathrm{NE} \equiv \tilde{\Omega}_\mathrm{NE} + \frac{1}{2}\int d\br\, \frac{\mu_N}{T}\rho_N + \frac{1}{2Z^2}\int d\br\, \frac{\mu_Q}{T}\rho_Q$. }:
\begin{equation}
\begin{split}
    \frac{\delta S}{\delta \rho_Q} = -\frac{1}{2Z^2} \frac{\mu_Q}{T}; & \quad \frac{\delta S}{\delta \rho_N} = -\frac{1}{2} \frac{\mu_N}{T};  \quad \frac{\delta S}{\delta e} = \frac{1}{T}.\label{eq:Euler-Lagrange-full}
\end{split}
\end{equation}
In this work, we assume that the temperature distribution is externally defined, so that $T(\br)$ is given as an input variable and is not a functional of the densities. This allows us to promptly simplify the first two Euler-Largrange equations and write the stationary conditions for the charge and number densities in terms of the dimensionless free-energy~$\tilde{F}_\mathrm{NE}$:
\begin{equation}\label{eq:euler}
    \frac{\delta \tilde{F}_\text{NE}}{\delta \rho_Q} = \frac{1}{2Z^2} \frac{\mu_Q}{T};\qquad  \frac{\delta \tilde{F}_\text{NE}}{\delta \rho_N} = \frac{1}{2} \frac{\mu_N}{T}\, .
\end{equation}

\subsection{Asymptotic functional approximation}
{In order to provide a closed functional form to Eq.~\eqref{eq:euler}, we rely on the hypothesis of \textit{local equilibrium} and define $\tilde{F}_\text{NE} \equiv \int d\br f(\br)/T(\br)$, where $f(\br)$ is the equilibrium free-energy density of the system at the local temperature~$T(\br)$. The problem is then reduced to the derivation of an approximation of $f(\br)$ suitable to describe the screening properties of the ionic fluid.}
The ideal-gas contribution to $f(\br)$ can be written exactly as $f_\text{id}(\br)=\sum_iT(\br)\rho_i(\br)\left[\ln\left(\Lambda^3(\br)\rho_i(\br)\right)-1\right]$, with $\Lambda(\br)$ the local de Broglie wavelength and $\rho_i(\br)$ the ionic density distributions for $i\in \{{+,-}\}$. To derive an approximation of the excess contribution, we start by separating a mean-field (Hartree-like) electrostatic term that expresses the self-interaction of the charge density  with itself, i.e., $f_\mathrm{H}(\br)=\frac{1}{2} \rho_Q(\br)\phi_\mathrm{H}(\br)$, with $\phi_\mathrm{H}(\br) = \int d\br' \rho_Q(\br')/|\br'-\br|$ the Hartree potential. The remaining (neutralized) part of the excess free-energy density $f^\text{exc}(\br)$ contains information about the short-range correlations in the fluid and its functional dependence from $\rho_{+/-}(\br)$ is generally unknown. However, in the limit of small and slowly varying inhomogeneities, which are implied by the asymptotic response of the ions' densities far from an external perturbation, a rigorous expression for $f^\text{exc}(\br)$ can be provided that is grounded on an asymptotic gradient expansion~\cite{Fleming1976,Senatore1980}. In particular, we rely on a square-gradient approximation (SGA) already adopted in the {non-isothermal treatment of liquid-vapor interfaces}~\cite{Bedeaux2003,Magnanelli2014}:
\begin{equation}\label{eq:gradient-approx}
 f^\text{exc}_\text{SGA}(\br) = f^\text{exc}_\text{LDA}(\br) +  \frac{1}{2}\sum_{ij} A^T_{ij} \nabla \rho_i(\br)\cdot \nabla \rho_{j}(\br)\, .
\end{equation}
Here, $A^T_{ij}$ are the temperature-dependent square-gradient coefficients{, defined as response functions of the homogeneous and isotropic system at the equilibrium temperature $T\equiv T(\br)$ and at some given reference densities, $\bar{\rho}_i$, which are set as input parameters.} In analogy with pioneering studies on the inhomogeneous electron gas~\cite{Fermi,Weizscker1935}, the first order term $f^\text{exc}_\text{LDA}(\br)$  of the gradient expansion is  defined as the local density approximation (LDA) of the excess free-energy density, which, in our case, also depends on the local temperature~$T(\br)$. Upon including the ideal-gas term $f_\text{id}(\br)$, local by definition, the total LDA contribution to $f(\br)$ can be expressed as a second-order expansion about the electroneutral homogeneous system at the local temperature~$T(\br)$ and reference densities $\bar{\rho}_i$:
\begin{equation}\begin{split}
    f_\text{LDA}(\br) \approx f^T({\{\bar{\rho}\}})&+\sum_i \mu^T_i \Delta\rho_i(\br) \\&+ \frac{1}{2}\sum_{ij} B^T_{ij} \Delta\rho_i(\br)\Delta\rho_j(\br)\, ,
\end{split}\end{equation}
{where $f^T({\{\bar{\rho}\}})$, $\mu^T_i=\tfrac{\partial f_\mathrm{LDA}}{\partial\rho_i} \big|_{\{\bar{\rho}\}}$ and $B^T_{ij} \equiv \tfrac{\partial^2  f_\mathrm{LDA}}{\partial\rho_i\partial \rho_j} \big|_{\{\bar{\rho}\}}$ are the free-energy density, the chemical potentials and the quadratic LDA coefficients of the homogeneous reference system~\cite{Ballone1981second}, respectively, evaluated at the local temperature $T=T(\br)$. 
Finally, we consider the interaction of the fluid charge density with the electrostatic potential $\phi_\text{ext}$ generated by an external charge distribution, i.e., $f_\text{ext}(\br) =\rho_Q(\br) \phi_\text{ext}(\br)$. Putting everything together and recasting the problem in the charge and total number density variables, $\rho_Q$ and $\rho_N$, the final asymptotic approximation of the free-energy density reads as follows:
\begin{equation}\begin{split}\label{eq:free-energy-density}
    f(\br) &\approx\ f^T({\{\bar{\rho}\}}) + \frac{1}{2} \mu^T_N \Delta\rho_N(\br) + \frac{1}{2Z^2} \mu^T_Q \rho_Q(\br)\\&+ \frac{1}{8}\left[B^T_{NN} \Delta\rho^2_N(\br) +A^T_{NN}\left|\nabla \rho_N(\br)\right|^2\right]\\&+\frac{1}{8Z^2}\left[ B^T_{QQ}  \rho^2_Q(\br)+   A^T_{QQ} \left|\nabla \rho_Q(\br)\right|^2\right] \\
    &+ \frac{1}{4Z}\left[B^T_{NQ}\rho_Q(\br) \Delta\rho_N(\br)  + A^T_{NQ}\nabla\rho_N(\br)\cdot\nabla\rho_Q(\br)  \right] \\
    &+ f_\mathrm{H}(\br) + f_\mathrm{ext}(\br) ,
\end{split}\end{equation}
where we defined the number-number, charge-charge, and charge-number coefficients as $X_{NN}=\sum_{ij} X_{ij}$, $X_{QQ} \equiv X_{++}+X_{--}-2X_{+-}$, and $X_{QN} = X_{NQ} \equiv X_{++}-X_{--}$, respectively, with $X_{ij} = B^T_{ij}$ or $A^T_{ij}$.  It is worth noticing that, from the underlying local equilibrium hypothesis, all quadratic LDA and SGA coefficients are mapped to equilibrium correlation functions at the local temperature $T(\br)$. In particular, $A^T_{ij}$ and $B^T_{ij}$ are defined from the fourth- and second-order momenta of the short-range part of the direct correlation function of the fluid at the local reference state $(T(\br),\bar{\rho}_i)$, i.e., $\tilde{c}^T_{ij}(r) = c^T_{ij}(r)+\beta Z_iZ_j/r$~\cite{Fleming1976}. In turn, this implies that $\tilde{c}^T_{ij}(r)$ is assumed to decay much faster than the variation rate of the temperature distribution~\cite{Vrugt2020}. A full derivation of Eq.~\eqref{eq:free-energy-density}, together with an explicit definition of the parameters $A^T_{ij}$ and $B^T_{ij}$, is reported in the Supplemental Material, Ref.~\onlinecite{suppmat}.}

\subsection{Steady-state condition for the charge density}

{Having obtained an explicit asymptotic approximation of $f(\br)$, we can now apply the steady-state Euler-Lagrange equations for the charge and number densities, as derived in Eq.~\eqref{eq:euler}. For simplicity, we will work under the hypothesis of a symmetric ionic fluid, so that the two ions differ from each other only by their electric charge. Under this assumption, $\mu_N$ is not a functional of $\rho_Q$ and all charge-number coefficients in Eq.~\eqref{eq:free-energy-density} are identically zero, i.e., $X_{NQ}=X_{QN}=0$.
This allows us to decouple the problem of Eq.\eqref{eq:euler}, so that independent solutions for $\rho_Q(\br)$ and $\rho_N(\br)$ can be found. In particular, we will only focus on the charge density decay, which is responsible for the electrostatic screening properties of the ionic fluid. }

{The aforementioned symmetry assumption carries strong implications on the properties of the charge chemical potential, $\mu_Q$. To see this, let us consider the non-equilibrium phenomenological equation for the charge flux~\cite{mazur1984}:
\begin{equation}\begin{split}
    \bm{j}_Q = 
    L_{qQ} \nabla \left( \frac{1}{T} \right) 
    - \frac{1}{T} \left(L_{QN} \nabla_T \mu_N + L_{QQ} \nabla_T \mu_Q\right)\, .
\end{split}\end{equation}
Here, $L_{qQ}$ is the Onsager coefficient that couples the charge flux with the heat flux $q$ \footnote{{Notice that the heat flux $\bm{j}_q = \bm{j}_e - \sum_i h_i \bm{j}_i$, where $h_i$ is the partial enthalpy of species $i$, and $\bm{j}_e$, $\bm{j}_i$ are the fluxes of energy and particles of species $i$, respectively, must be considered when we address $L_{qQ}$ in a manner that \textit{i)} is independent of the zero of the energies, and \textit{ii)} allows to single out the \textit{isothermal} gradient of the chemical potential, see Ref. \cite{Bertossa2019}. Also notice that in Ref. \cite{mazur1984} the notation $L_{q'z}$ is used for $L_{qQ}$.}}, $L_{NQ}$ is the Onsager coefficients between the fluxes of charge and total number of particles, and $L_{QQ}$ is the charge-charge Onsager coefficient. Then, the symmetry of the ionic fluid implies that we can neglect the contribution to $\bm{j}_Q$ due to thermoelectric effects, $L_{qQ}=0$, as well as the coupling between the charge and total number of particles, $L_{QN}=0$. If no particle exchange is allowed through the system, we can further assume that the stationary state is characterized by a vanishing charge flux, i.e., $\bm{j}_Q=0$. Therefore, the \textit{isothermal} gradient of $\mu_Q$ is everywhere zero, implying that the charge chemical potential depends on the spatial coordinates only through the variations of the local temperature,  $\mu_{Q}(\br)\equiv\mu^T_{Q}$. This result is particularly important, as it allows us to rewrite the Euler-Lagrange equation for the charge density as
}

{
\begin{equation}\label{eq:EL-simplified}
   \frac{\delta F_\text{NE}}{\delta \rho_Q(\br)}=\frac{1}{2Z^2} \mu^T_Q 
\end{equation}
where $F_\text{NE} = \int d\br f(\br)$ represents the non-equilibrium free-energy defined as the integral of the local-equilibrium free-energy density. In fact, the so derived steady-state condition is formally equivalent to the usual Euler-Lagrange equation associated with the free-energy functional minimization at thermal equilibrium. In this context, $\mu^T_Q$ acts as a Lagrange multiplier for enforcing the stationary nature of the charge chemical potential with respect to isothermal variations, i.e., $\nabla_T\mu_{Q}(\br)=0$. At a more practical level, this result allows us to simplify the r.h.s. of Eq.~\eqref{eq:EL-simplified} with the analogous term coming from the free-energy density approximation of Eq.~\eqref{eq:free-energy-density}. Crucially, this implies that the only linear term in the asymptotic expansion of $f(\br)$ comes from the interaction of $\rho_Q(\br)$ with the external perturbation, thus guaranteeing that, for positive values of the quadratic expansion coefficients, the constrained functional is bounded from below.}

To obtain an explicit solution to Eq.~\eqref{eq:EL-simplified}, we assume that both the external perturbation $\phi_\text{ext}(\br)$ and the temperature profile $T(\br)$ have planar symmetry, so that we will only refer to one-dimensional variations along~$x$. 
{The resulting steady-state condition consists in the following  integral-differential equation:} 
\begin{equation}
    \label{eq:local-eq}\begin{split}
     \frac{1}{4Z^2}\left\{B^T_{QQ} \rho_Q(x) - \frac{d}{dx}\left[A^T_{QQ}\, \rho'_Q(x)\right] \right\} + \phi(x)=0\, ,
\end{split}
\end{equation}
{where $\phi = \phi_\mathrm{H} + \phi_\mathrm{ext}$ is the total electrostatic potential. To get rid of the integral dependence of the electrostatic potential from $\rho_Q$ in Eq.\eqref{eq:local-eq}, we can now take advantage of the Poisson equation, i.e., $\phi''(x)= -4\pi[\rho_Q(x)+\rho_\mathrm{ext}(x)]$. In particular, we consider the screening behavior of the ionic fluid away from the external perturbation, where we can set $\rho_\mathrm{ext}(x)=0$. This leaves us with a homogeneous differential equation which describes the asymptotic decay of the charge density under steady-state conditions:}
\be\label{eq:master}\begin{split}
0 = -16\pi Z^2 \rho_Q(x) &+2 \tfrac{\partial B^{T}_{QQ}}{\partial x}\,\rho'_Q(x) + B^T_{QQ}\, \rho''_Q(x) \\& -3 \tfrac{\partial A^{T}_{QQ}}{\partial x}\,\rho'''_Q(x)
- A^T_{QQ}\, \rho''''_Q(x)\, .
\end{split}\ee 
{In what follows, we consider that the external potential is originated by a planar charged wall of surface charge density $\eta$, i.e., $\phi_\text{ext}(x) = -\eta x / (4\pi)$. Assuming perfect screening, this implies that $\int_{0}^\infty dx\, \rho_Q(x)=-\eta$, providing a boundary condition to the solution of Eq.~\eqref{eq:master}.}

\subsection{Thermal equilibrium limit}

From Eq.~\eqref{eq:master}, we immediately recover the equilibrium limit of vanishing temperature gradients. In this case, the coefficients $A_{QQ}$ and $B_{QQ}$ are independent from the spatial coordinates and the solutions  can be searched in the form of exponentials with complex arguments, $e^{\kappa x}$. The four decay factors $\kappa$ are the solutions of
\begin{equation}\label{eq:equilibrium}
    \kappa^2 = \frac{B_{QQ}\pm \sqrt{B_{QQ}^2-64\pi Z^2 A_{QQ}}}{2A_{QQ}}\, .
\end{equation}
Their complex nature reflects the possibility of obtaining either a monotonic or an oscillatory decay of the charge density, depending on the reference thermodynamic state. In fact,
the locus of points in the phase diagram that determines a transition between a monotonic and oscillatory behavior (Fisher-Widom line~\cite{fisher-widom}) is found for vanishing discriminants $B_{QQ}^2-64\pi Z^2A_{QQ}=0$. Crucially, this result is entirely equivalent to what originally derived in Refs.~\cite{Ballone1981first,Ballone1981second} for the charge density of a classical plasma, as well as to what obtained by other means in the study of the screening lengths of bicomponent ionic fluids~\cite{carvalho1994}.

\section{Discussion}

\begin{figure}[t!]
    \centering
    \includegraphics[width=0.7\columnwidth]{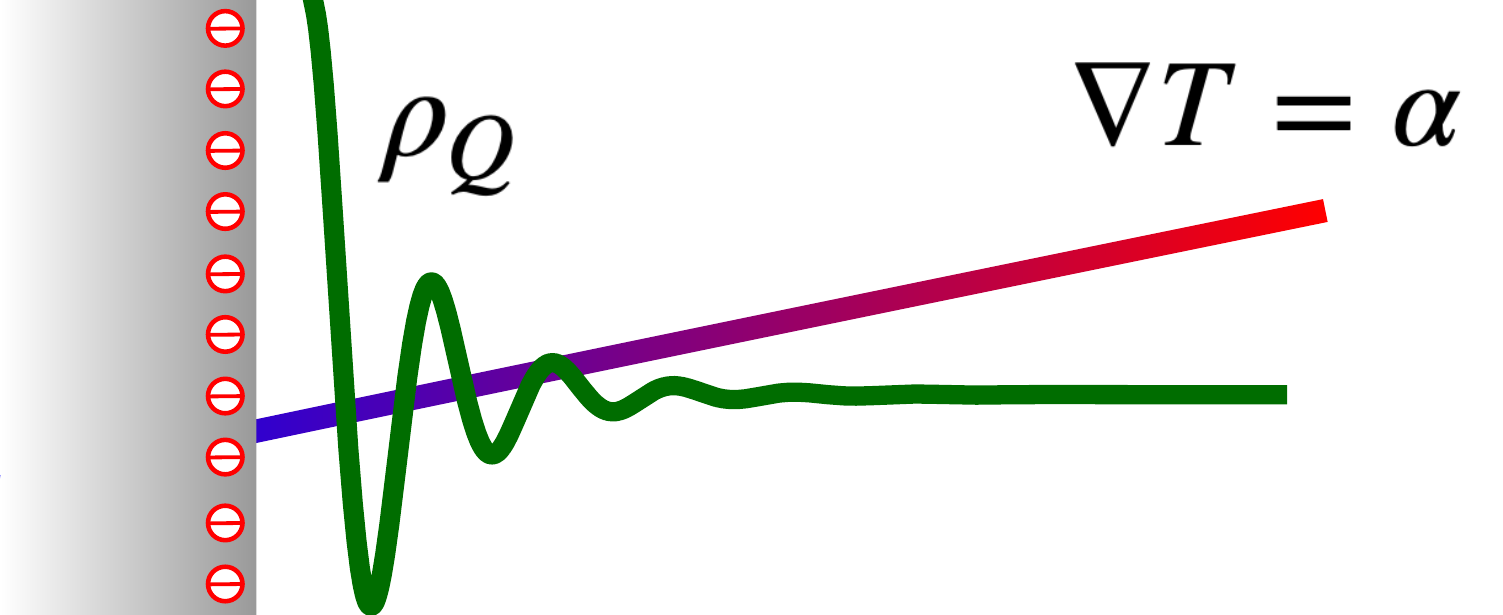}
    \caption{
    {Sketch of the physical system studied in this work: an external surface charge is screened by the ionic charge density $\rho_Q(x)$ in the presence of a thermal gradient $\alpha$. %
    }
    }
    \label{fig:wkb}
\end{figure}

Because of the spatial dependence of the local-density and gradient coefficients brought by the temperature variations, finding an analytical solution to Eq.~\eqref{eq:master} is in general a hard problem. Here, we will limit our discussion to a linear temperature profile along $x$, i.e., $T(x)=T^0+\alpha x$, with $T^0$ a reference temperature and $\alpha$ the temperature gradient (Fig.~\ref{fig:wkb}). In turn, this allows us to linearize the coefficients $A^T_{QQ}$ and $B^T_{QQ}$ about~$T^0$, i.e.,  $A^T_{QQ}\approx A^0_{QQ}+\tfrac{\partial A_{QQ}}{\partial T}\big|_{0}\alpha x$ and $B^T_{QQ}\approx B^0_{QQ}+\tfrac{\partial B_{QQ}}{\partial T}\big|_{0}\alpha x$. 
{In this work, the (neutralized) direct correlation functions that enter the calculation of the reference coefficients $A^0_{QQ}$ and $B^0_{QQ}$, together with their derivatives,} are computed using the hypernetted chain (HNC) approximation~\cite{Hansen_McDonald}, which is known to provide a particularly good description of ionic fluids outside the region of the phase diagram delimited by the spinodal line~\cite{Abernethy1980,Belloni1993,Johan1993}.

\subsection{WKB method}\label{sec:WKB}

{Before proceeding to discuss the specific solutions to our problem, it is worth pointing out that the equilibrium charge density profiles at the reference temperature $T_0$ are expected to decay much faster than the applied linear temperature distribution. In more rigorous terms, we expect that the characteristic screening lengths $\lambda$ that can be derived from the equilibrium solutions of Eq.~\eqref{eq:equilibrium} satisfy the inequality $\lambda \ll T^0/\alpha$. }
This limit behavior allows us to recognize a formal analogy with the semiclassical regime of quantum mechanics, where the particle's wavefunction~$\psi$ displays quick variations over the action range of the potential~$V$. In particular, under the formal maps $\psi \mapsto \rho_Q$ and $V\mapsto T$, it must be possible to interpret the temperature gradient~$\alpha$ as the small parameter of an asymptotic expansion that mimics the semiclassical solution of the Schr\"odinger equation in the limit of small~$\hbar$. {Therefore, we expect that the solutions to Eq.~\eqref{eq:master} can be directly obtained using the WKB~method~\cite{bender78}. In order to set up the WKB routine,} we can start applying the formal substitution $\rho_Q(y) = \exp\left[S(y)\right]$, where $S(y)$ is related to the potential of mean force of the charge density written as a function of the reduced variable $y\equiv\alpha x$.
Then, the WKB method reads as follows~\cite{bender78}: \textit{i)}~consider the asymptotic expansion of $S(y)$ up to order $\alpha^0$, i.e., $S(y) \approx \frac{1}{\alpha} S_0(y) + S_1(y)$; \textit{ii)}~expand the {charge-density differential} equation in terms of $\alpha$ and retain only the dominant terms; \textit{iii)}~find the first order term~$S_0$; \textit{iv)}~find $S_1$ recursively. {In the next subsections we show how this procedure can be successfully applied to find LDA and SGA asymptotic solutions.}

\subsection{LDA solutions}\label{sec:LDA}
We start our discussion by retaining only the first line of Eq.~\eqref{eq:master}, corresponding to the LDA screening regime. In this case, exact solutions exist in terms of linear combination of $z^{-1/2} I_1 \left(2\sqrt{z}\right)$ and $z^{-1/2} K_1 \left(2\sqrt{z}\right)$, where $I_1$ and $K_1$ are the order-1 modified Bessel functions of first and second kind, respectively, 
and $z \equiv 16\pi Z^2 \left(B^0_{QQ} + \tfrac{\partial B_{QQ}}{\partial T}\big|_{0} \alpha x \right)\big(\tfrac{\partial B_{QQ}}{\partial T}\big|_{0} \alpha\big)^{-2}.$
At large~$z$, both $I_1$ and $K_1$ can be approximated in terms of their asymptotic forms,  giving the LDA charge-density profile 
\begin{equation}\label{eq:pure-charge-solution-asymp-LDA}
    \rho_Q(z) \sim z^{-3/4}\, \left(C_+ \, e^{+2\sqrt{z}} + C_- \,  e^{-2\sqrt{z}}\right)\, ,
\end{equation}
where the constants $C_{\pm}$ are found from the boundary conditions. When considering finite values of $x$, the solution so derived is expected to hold {for sufficiently small values of $\alpha$.} For example, in the simplest case of considering the DH limit of vanishing short-range interactions between the ions, Eq.~\eqref{eq:pure-charge-solution-asymp-LDA} is found to apply whenever $\lambda_D\ll T^0/\alpha$, with $\lambda_D$ the Debye screening length. 
In fact, we find that the solution of Eq.~\eqref{eq:pure-charge-solution-asymp-LDA} can be directly recovered using the WKB~method described in Sec. \ref{sec:WKB}.
{In particular, after the LDA equation is expanded in terms of $\alpha$ and only the dominant terms are retained,} the final result is $S_0 = \pm 2\alpha\sqrt{z} + C'_{\pm}$ and $S_1 = -\tfrac{3}{4} \ln(z) +C'$, with $C'_{\pm}$ and $C'$ integration constants, which leads to the same asymptotic form of the exact LDA solution obtained in~Eq.~\eqref{eq:pure-charge-solution-asymp-LDA}. We redirect the reader to \cite{suppmat} for the explicit details of the calculation. {We note that the charge integral of Eq.~\eqref{eq:pure-charge-solution-asymp-LDA} tends to a constant $Q_0$ faster than the variation rate of the space dimensionality $x$. When considering the decaying solution for $x\to\infty$, for instance, we get 
\begin{equation}
    Q(x) \sim C_Q\, x^{-\frac{1}{4}}\exp\left(-2\sqrt{\frac{16\pi Z^2}{\tfrac{\partial B_{QQ}}{\partial T}\big|_{0}\alpha}x}\right) + Q_0\, .
\end{equation}
This result is particularly important, as it guarantees the capability of the fluid to 
perfectly screen an external charge over a finite distance. In turn, this justifies the adoption of the boundary condition $Q_0=-\eta$ as a physical criterion to determine the  free parameters of Eq.~\eqref{eq:pure-charge-solution-asymp-LDA}\footnote{Notice that the $x$-integral of the charge density, $\rho(x)$, has dimensions of a surface density, i.e., $[Q(x)] = e/\sigma^2$.}.}

\begin{figure}[t!]
    \centering
    \includegraphics[width=0.95\columnwidth]{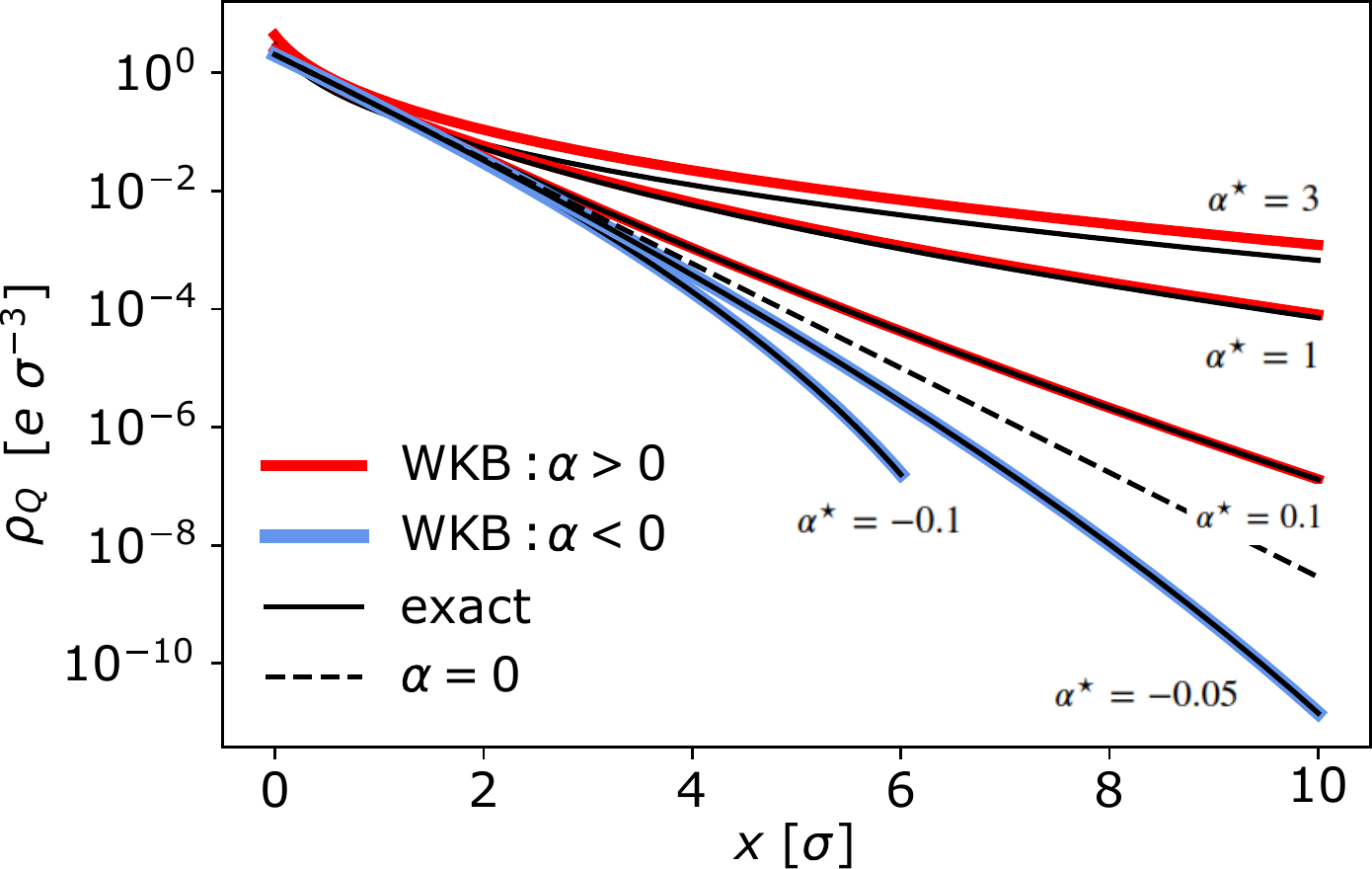}
    \caption{Asymptotic decay of the LDA charge density of
a symmetric ionic fluid screening a negative surface charge
density $\eta=-1.0\, e/\sigma^2$ while subject to a temperature gradient
$\alpha=\alpha^\star T^0/\sigma$, with $\alpha^\star$ indicated in the plot. Red (blue) solid
lines: WKB solutions for positive (negative) $\alpha$. Black solid
lines: exact solutions. Black dashed line: equilibrium solution.}
    \label{fig:lda_comparison}
\end{figure}

To test the accuracy of our solutions, we consider a symmetric molten salt of unit valence $\pm1$ \footnote{{For any static configuration, no evidence is found for integer charges, which are manifested in transport processes only \cite{Pendry1984, Pegolo2020,Resta2021}. Nonetheless, we used integer charges to be consistent with established literature \cite{Wurger2020,Kjelstrup2023}, where the same (integer) charge is used in both the electrostatic interaction and the charge flux. We believe that this holds true especially for diluted electrolytes.}}, where the short-range interaction between the ions is given by a single Lennard-Jones (LJ) potential. Under this choice, the LJ parameters $\sigma$ and $\varepsilon$ are taken as units of length and energy, respectively. Given the hypothesis of planar symmetry, we will consider all throughout the screening of a uniform surface charge density $\eta=-1.0\, e/\sigma^2$, while the linear temperature profile $T(x)=T^0+\alpha x$ is defined by setting the origin $x=0$ at the position of the planar charged surface. We choose a reference thermodynamic state given by a mean number density  $\bar{\rho}_N=0.5\,  \text{ions}/\sigma^3$ and temperature $T^0=2000\, \varepsilon/k_B$. In agreement with the thermal equilibrium solutions provided in Eq.~\eqref{eq:equilibrium}, the large value of $T^0$ is expected to yield a monotonic decay of $\rho_Q$ which can be well described by the LDA screening regime.  
Fig.~\ref{fig:lda_comparison} reports the asymptotic behavior of the charge-density profiles computed at various temperature  gradients $\alpha$. When compared with the equilibrium case of $\alpha=0$, our solutions are capable to reproduce the expected decrease in the effectiveness of the electrostatic screening at increasing temperatures ($\alpha>0$), a phenomenon that becomes particularly pronounced at large values of $\alpha$. Conversely, the screening is enhanced when approaching cooler regions ($\alpha<0$). The WKB solution is found in perfect agreement with the exact charge-density decay up to temperature gradients as large as $\alpha = 1.0\, T^0/\sigma$. This result remarks the applicability of the asymptotic approximation in describing steady-state regimes that are strongly driven away from thermodynamic equilibrium. It should be noted that the solution reaches a turning point at $B^T_{QQ}=0$, under which the charge density shows an algebraically damped oscillation similar to what observed in quantum-mechanical semiclassical solutions. In our case, however, this behavior identifies a breakdown of LDA at low temperatures, implying that better functional approximations are required. 

\subsection{SGA solutions}\label{sec:SGA}
While LDA is generally capable of reproducing the monotonic decay of $\rho_Q$, an oscillatory screening regime characteristic of low temperatures can only be described by solving the full SGA equation reported in Eq.~\eqref{eq:master}. In this case, exact solutions cannot be found and we are forced to rely on the WKB approximation straight away. Following the procedure described Sec. \ref{sec:WKB}, we then look once again for the asymptotic expansion at small $\alpha$ of the potential of mean force in the form of $S \approx \frac{1}{\alpha} S_0 + S_1$, which defines the charge density as $\rho_Q\propto \exp[S]$.
After a tedious but straightforward calculation, four solutions for $S_0$ and a pair of solutions for $S_1$ are obtained; their explicit functional forms are reported in \cite{suppmat}. Here again, $S_0$ is associated with dominant, exponential-like terms, while $S_1$ contains logarithmic terms that account for algebraic contributions to the total charge-density decay. Note that when taking the equilibrium limit of $\alpha \to 0$, $S_1$ can be neglected, and the potential of mean force reduces to $S = \frac{1}{\alpha}S_0 = \kappa x$, where $\kappa$ corresponds to the pure exponential decay factors reported in Eq.~\eqref{eq:equilibrium}.

\begin{figure}[t!]
    \centering
    \includegraphics[width=0.95\columnwidth]{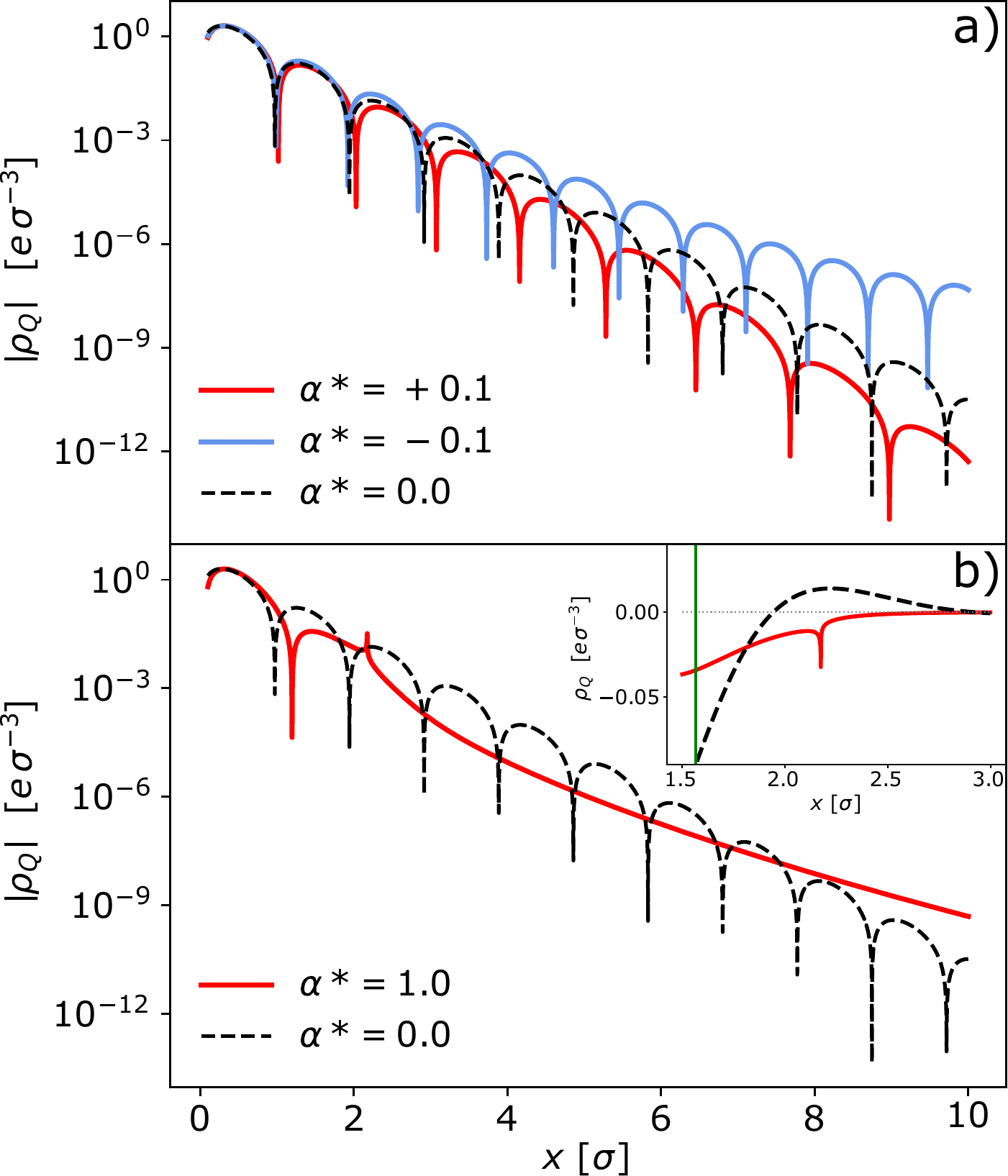}
    \caption{a) WKB asymptotic decay of the SGA charge density of a symmetric ionic fluid subject to positive and negative temperature gradients $\alpha=\alpha^\star T^0/\sigma$. b) Crossover between oscillatory and monotonic regime of the charge density decay for a positive temperature gradient $\alpha=1.0\, T^0/\sigma$. {In both cases, the screening of a negative surface charge density of $\eta=-1.0\, e/\sigma^2$ is considered.} The absolute value $|\rho_Q|$ is taken to report the profiles on a logarithmic scale. Inset: normal scale zoom around the crossover point of $\rho_Q$; the vertical green line indicates the position corresponding to the Fisher-Widom temperature at equilibrium.}
    \label{fig:figure-2}
\end{figure}

In Fig.~\ref{fig:figure-2}-a) we show the WKB asymptotic profiles computed at $\alpha = \pm 0.1\, T^0/\sigma$. This time, the reference thermodynamic state is defined by a temperature  $T^0 = 500\, \varepsilon/k_\text{B}$, for which the equilibrium system is in a highly oscillatory regime. Unlike the LDA case, we observe that the screening becomes increasingly effective when going towards warmer regions of the fluid, with oscillation wavelengths that are progressively stretched out. On the other hand, an underscreening typical of concentrated ionic fluids is found when moving towards lower temperatures, which comes together with shorter and shorter oscillations. Note that, in contrast to LDA, the SGA solutions remain valid down to $T=0$, i.e., $x=10\, \sigma$, for which the system tends to display a crystal-like behavior of pure oscillatory charge-density variations.  It is also possible to test our solutions  over temperature windows that embrace the transition between monotonic and oscillatory regimes. 
{In particular, under the hypothesis of local thermodynamic equilibrium, we expect that the same structural transition observed when moving across the Fisher-Widom line in the equilibrium phase diagram is also observed when overcoming the crossover temperature in our physical system.}
This is shown in Fig.~\ref{fig:figure-2}-b), where a {large} temperature gradient is applied starting from the same reference temperature~$T^0$. In this case, a sharp crossover towards a monotonic screening regime is observed  when crossing a critical temperature~$T^*$ associated with a given distance $x^*$.  As reported in the inset, the structural crossover is identified by a characteristic cusp behavior in the charge-density profile.  On this regard, it is worth noticing that the temperature $T^*$ is close to, although not corresponding to, that associated with the Fisher-Widom line of the equilibrium system at the reference density $\bar{\rho}_N$ (green line). {This discrepancy is related to the fact that, for this extreme case, the square-gradient parameter $A_{QQ}^T$ would be better approximated by including higher orders in the Taylor expansion around $T^0$. A more detailed discussion on this matter is reported in~\cite{suppmat}.}  An opposite crossover for $\alpha<0$, from monotonic to oscillatory, can be similarly obtained when starting from a large value of $T^0$.

\subsection{{Quantitative model for dilute electrolytes}}

We now provide a thorough quantitative analysis of our effect by taking as an example the electrostatic screening of a charged surface in aqueous solutions of NaCl and NaOH, in the presence of a thermal gradient. We assume a surface charge density of $\eta=1 e/\mathrm{nm}^2$ and ionic concentration of $\bar{\rho}_N = 0.01$~M
\footnote{{Despite such a low $\bar{\rho}_N$, steric interactions between the ions may become substantial close to the charged surface \cite{Kilic2007}. Nonetheless, a detailed study of the behavior of $\rho_Q$ close to the surface is beyond the scope of the present example, so that we neglect steric effects.}}.
Consistently with previous works~\cite{Dhur2006,Ly2018}, we then apply a negative temperature gradient of $\alpha=-10$~K/$\mu$m, starting from a reference temperature of $T^0=300$~K at the surface position.
The dielectric permittivity of water is fixed at $\varepsilon=80$, which accounts for scaling the charges by $\varepsilon^{-1/2}$ all throughout in our equations. Given the low ionic concentration, the effect of short-range ionic interactions can be safely neglected and the charge-density decay can be simply described by taking the dilute Debye-H\"uckel limit of the LDA solutions. This corresponds to setting $B^T_{QQ}=4T(x)/\bar{\rho}_N$~in~Eq.~\eqref{eq:pure-charge-solution-asymp-LDA}. Note that by neglecting any ionic interaction beyond the Coulomb pair-potential, the NaCl and NaOH charge density profiles are expected to be identical in this example.

\begin{figure}[t!]
    \centering
    \includegraphics[width=0.95\columnwidth]{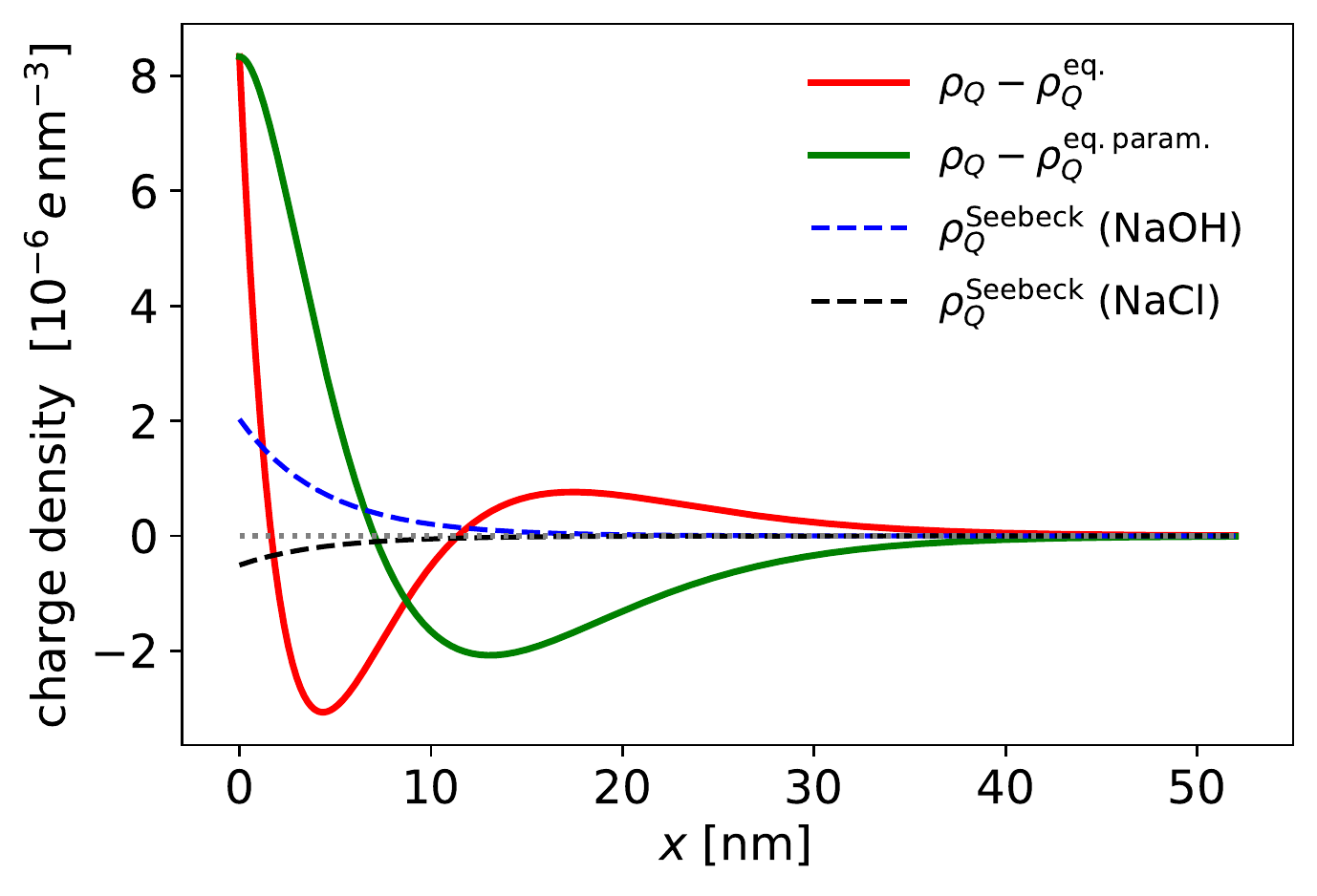}
    \caption{{
    Effect of a negative thermal gradient on the charge distribution of a typical dilute electrolyte. Deviations of our solutions from the equilibrium distribution (red line), and from a $T$-dependent parametrization of the equilibrium distribution (green line) are shown as a function of the distance $x$ from the charged wall of surface charge density $\eta = 1e/\mathrm{nm}^2$. The Seebeck charge distribution  represented through a DH model for 0.01~M aqueous solutions of NaOH (dashed blue line) and NaCl (dashed black line) are also reported.}
    }
    \label{fig:Seebeck}
\end{figure}

In Fig.~\ref{fig:Seebeck} we report the deviation of our solution, Eq.~\eqref{eq:pure-charge-solution-asymp-LDA}, from the equilibrium DH solution at $T^0$, 
\begin{equation}\label{eq:DH}
    \rho^\text{eq.}_Q(x) = - \eta \, \kappa_D^0\,  e^{-\kappa_D^0\, x}\, ,
\end{equation}
as well as from a $T$-dependent parametrization of the equilibrium DH solution:
\begin{equation}\label{eq:DH-param}
    \rho_Q^\text{eq.param.}(x) = - \eta \frac{\left(\kappa_D^T\right)^2}{\kappa_D^0} e^{-\kappa_D^T\, x}\, ,
\end{equation}
where we defined $\kappa^T_D\equiv\sqrt{4\pi e^2 \bar{\rho}_N /[\varepsilon T(x)]}$ as the $T$-dependent Debye decay factor. {At $T=T^0$, this results in a DH decay length $(\kappa_D^0)^{-1} \approx 4.36$nm.}
As shown in the Figure, the expected increase in the screening effectiveness with respect to the equilibrium solution of Eq.~\eqref{eq:DH} is reflected in an overall accumulation of negative charge close to the charged surface. An opposite behavior is instead observed when considering the deviation with respect to the parametrized solution of Eq.~\eqref{eq:DH-param}, where the strengthening of the screening generated by the temperature gradient tends to be overestimated. Importantly, however, deviations of similar magnitude are observed in both cases, 
demonstrating that our WKB solutions can be used to obtain qualitatively different results than a mere $T$-dependent parametrization of the equilibrium solutions. 

To complement the previous discussion, we provide an estimate of the relative importance of our effect with respect to the expected charge accumulation generated by the ionic Seebeck effect. In fact, while the presented theory neglects thermodiffusion phenomena, the different thermal diffusion of the two ions in solution is expected to give rise to a macrosopic electric field through the system associated with the accumulation of opposite charge at the two walls of an hypothetical thermoelectric cell. The Seebeck electric field is defined as $E_S = \chi \alpha$, with $\chi$ the Seebeck coefficient. Following Ref.~\cite{Ly2018}, we consider in particular $\chi=-0.2$ mV/K and $\chi=+0.05$ mV/K for NaOH and NaCl solutions, respectively. The functional form of the Seebeck charge density profile in electrolytes has been the subject of recent investigations \cite{Stout2017,Janssen2019}. A comparison {against the magnitude} of our {effect} can be carried out assuming a DH profile \cite{Ly2018}, which integrates to a surface charge density $\eta_S = \varepsilon \chi \alpha/(4\pi)$ compatible with the expected Seebeck field. Note that the Seebeck charge density is expected to be several orders magnitude smaller than the ionic charge density responsible for the screening of the surface charge, i.e., $\rho_Q^\mathrm{Seebeck}(x)/\rho_Q^\mathrm{eq.}(x) = \eta_S/\eta \sim 10^{-6}$. 
In Fig.~\ref{fig:Seebeck}, the so computed Seebeck profiles are reported against the deviations associated with our non-equilibrium solutions. We find that while the effect of thermodiffusion is not negligible, the predicted correction to the screening properties of the electrolyte plays a predominant role in describing the variations of ionic charge density induced by the application of a linear temperature gradient.
Importantly, this result appears to be largely independent of the temperature gradient. In fact, at the first order in $\alpha$, we have the following analytical expression for the deviation predicted by our solutions and the Seebeck charge density distribution:
\begin{equation}
   \frac{\rho_Q(x) - \rho_Q^\mathrm{eq.}(x)}{\rho_Q^\mathrm{Seebeck}(x)} \approx \frac{4\pi\eta}{4 T^0 \kappa_D^0 \varepsilon \chi} \left[
        1 - 3 \kappa_D^0 x + (\kappa_D^0 x)^2 
    \right]\, ,\label{eq:ratio}
\end{equation}
which we find to apply for a wide range of relatively small values of $\alpha$.

\subsection{{Potential drop between planar charged walls}}

\begin{figure}[t!]
    \centering
    \includegraphics[width=0.95\linewidth]{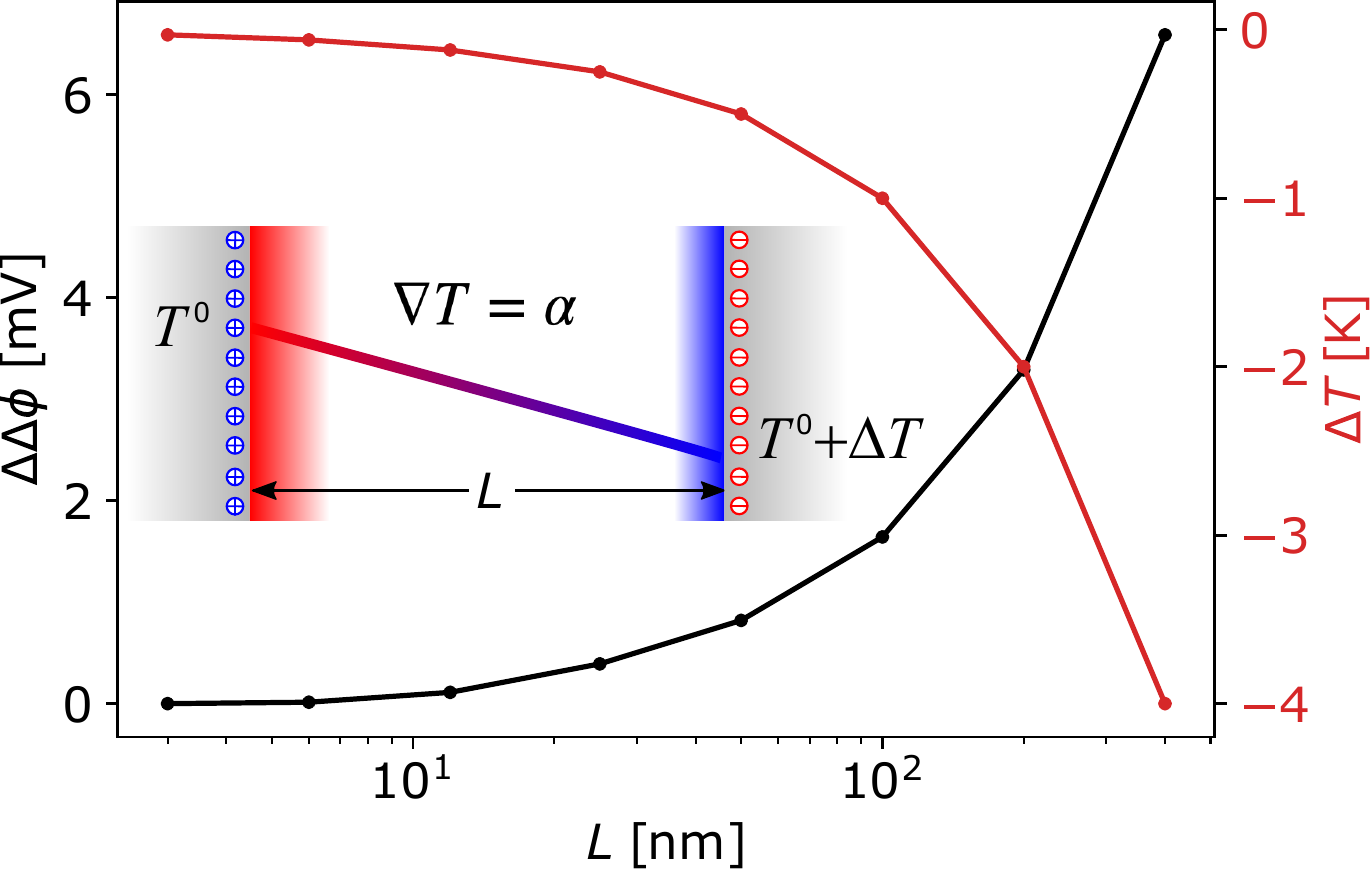}
    \caption{{Black line: effect of a stationary temperature gradient $\alpha=-10$~K/$\mu$m on the electrostatic potential drop of a dilute electrolyte solutions embedded between two oppositely charged walls, measured with respect to the equilibrium Debye-H\"uckel predictions. Red line: temperature difference between the charged walls. In both cases, results are shown as a function of the distance between the two walls, reported on a log-scale to zoom the behavior at small $L$. 
    Inset: sketch of the physical system under study.
   }
    }
    \label{fig:deltaV}
\end{figure}

    As a final application of our theory we consider how the steady-state electrolyte model previously introduced enters the description of the electrostatic potential drop between two oppositely charged walls at a given distance $L$ that are kept at a temperature difference $\Delta T$. Building on the previous example, we then fix the reference temperature $T^0$ at the position of the positively charged wall, while considering a constant temperature gradient $\alpha=-10$~K/$\mu$m through the fluid, so that the negatively charged wall is found at a temperature $T^0+\Delta T < T^0$, with $\Delta T=\alpha L$.  In Fig.~\ref{fig:deltaV}, we report the effect of the temperature gradient on the electrostatic potential difference between the two walls, as a function of $L$, in terms of the deviation $\Delta\Delta\phi$ from equilibrium solutions.  We find that the effect predicted by our non-equilibrium solutions increases linearly with the box size, going from 0.11mV for $L=12$nm to 6.6mV for $L=400$nm. The large nature of the electrostatic potential variations at large $L$ has to be related to the large deviations of the electrostatic potential profile in the proximity of the cooler wall at $T^0+\Delta T$. The absolute charge density profiles and the non-equilibrium electrostatic potential variations with respect to the reference DH model are reported in the Suppl.~Mat.~\cite{suppmat} for the various values of $L$ considered. Despite the predicted linear increase of our  effect with the system size, we note that no reciprocal interaction between the two walls is observed for $L>50$nm, i.e., when $L$ is much larger that the characteristic screening length of the fluid. This is clarified by the absolute electrostatic potential drop computed at the equilibrium temperature $T^0$, which, for $L>50$nm, shows a saturation at $\Delta \phi_\text{DH}\approx -2$V. {Note that this value is comparable to (the opposite of) twice the Gouy-Chapman single-surface  potential~\cite{Gouy1910,chapman1913}, $\Delta\phi_\text{GC} = \eta/(4\pi\epsilon\kappa_D^0) \approx 1$V.}
    Within this regime, separate non-equilibrium screening profiles computed at the reference temperature $T^0$ corresponding to $T(x=0)$ and $T(x=L)$ could in principle be used to independently represent the screening of the two surface charges, so that the problem can be once again formulated in terms of single interfaces. Conversely, we observe that within the domain of distances for which the two interfaces feel the reciprocal interaction, $L<50$nm, a single set of solutions for the ionic charge density must be adopted in order to properly describe the potential drop through the system. In particular, noticeable deviations of the electrolyte screening profiles with respect to the equilibrium DH solutions are found which are associated with electrostatic potential differences of the order of tens of mV, reaching  $\approx 0.9$mV at $L=50$nm. Note that the magnitude of this effect overshadows the voltage drop associated with the Seebeck effect at the same walls' distance, which, for the NaCl and NaOH solutions previously considered, accounts for just $-0.025$mV and $+0.1$mV, respectively. 

\section{Conclusions}\label{sec:conclusions}

The analytical study introduced in this work provides a rigorous framework to describe the screening properties of ionic fluids under a stationary temperature gradient. As a major result, we have shown that the striking parallelism with the semiclassical (WKB) regime of quantum-mechanics can be used as a powerful approach to extend an equilibrium theory to non-equilibrium steady-states. {The presented WKB method provides a precise protocol to obtain solutions under small thermal gradients, therefore avoiding a cascade of approximations which would be necessary at each step of the derivation whenever a standard linearization approach is followed~\cite{Dhont2008}. By relying on rigorous asymptotic approximations of the local free-energy density, our theory is capable to represent ionic fluids at arbitrary large ionic concentrations. This is reflected 
by the SGA charge-density profile, which allows us to predict the structural crossover between monotonic and oscillatory screening regimes induced by the temperature variation, as expected from the assumption of local thermodynamic equilibrium.} 

{In the paradigmatic case of dilute electrolyte solutions, the deviations of our charge-density predictions from thermal equilibrium have a larger amplitude than the charge distribution generated by the ionic Seebeck effect. Therefore, as long as the Seebeck effect can be experimentally probed, we expect that the magnitude of our effect should be similarly captured. This holds true, in particular, for measures of the thermal-gradient induced deviation from equilibrium of the potential difference between planar oppositely charge walls, which we find to overshadow the Seebeck voltage. 
Further experimental design may include the measurement of transients changes in the potential induced by laser heating of electrodes \cite{Climent2004, Liu2021}, modifications in the dynamics of charged colloids due to temperature gradients \cite{Dhont2008}, or a non-equilibrium extension of experiments that directly probe the electrolyte-mediated interaction between planar charged surfaces~\cite{Perkin2016}.}

Our theory can in principle be applied to any pair potential, such as those entering coarse-grained models of room temperature ionic liquids~\cite{salazar2020}. Moreover, beyond planar perturbations, the problem could be similarly formulated for external potentials and temperature profiles that have cylindrical and spherical symmetry. 
In addition, suitable extensions of the method could be derived to reproduce the interplay of ionic and dielectric screening, ubiquitous in aqueous electrolyte solutions~\cite{Coupette2018}, {thus overcoming the simplified solvent representation as a temperature-independent dielectric constant.}
Finally, we envision that our study could be used as a stepping stone for {providing a rigorous microscopic understanding of} the charge separation induced by temperature gradients when coupling the number and charge density in asymmetric ionic fluids, a phenomenon which is at the cornerstone of {a first-principles treatment of the ionic Seebeck effect.}

\begin{acknowledgements}
We thank Pietro Ballone for sharing the HNC code used to perform the calculations of the reference ionic system. 
AG acknowledges funding from the Swiss National Science Foundation.
FG acknowledges funding from the European Union's Horizon 2020 research and innovation programme under the Marie Sk\l{}odowska-Curie Action IF-EF-ST, grant agreement no. 101018557 (TRANQUIL).
\end{acknowledgements}

\end{document}